\documentclass[aps,prb,twocolumn,showpacs,groupedaddress]{revtex4}

\usepackage{graphicx}
\usepackage{latexsym}
\usepackage{amssymb}

\begin{document}

\title{Successive superconducting transitions and Anderson localization 
effect in Ta$_{2}$S$_{2}$C}

\author{J\"{u}rgen Walter}
\email[]{juergen.walter@chemiemetall.de}
\altaffiliation{CM Chemiemetall GmbH, Niels-Bohr-Str. 5, Bitterfeld, 
Germany}
\affiliation{Department of Materials Science and Processing, Graduate
School of Engineering, Osaka University, 2-1, Yamada-oka, Suita, 565-0879,
JAPAN}

\author{Itsuko S. Suzuki}
\email[]{itsuko@binghamton.edu}
\affiliation{ Department of Physics, State University of New York at
Binghamton, Binghamton, New York 13902-6016}

\author{Masatsugu Suzuki}
\email[]{suzuki@binghamton.edu}
\affiliation{ Department of Physics, State University of New York at
Binghamton, Binghamton, New York 13902-6016}


\date{\today}

\begin{abstract}
A complex carbide Ta$_{2}$S$_{2}$C consists of van der Waals (vdw)-bonded
layers with a stacking sequence $\cdots$ C-Ta-S-vdw-S-Ta-C- $\cdots$ along the
$c$ axis.  The magnetic properties of this compound have been studied from
DC and AC magnetic susceptibility.  Ta$_{2}$S$_{2}$C undergoes successive
superconducting transitions of a hierachical nature at $T_{cl} = 3.61 \pm
0.01$ K [$H_{c1}^{(l)}(0) = 28 \pm 2$ Oe and $H_{c2}^{(l)}(0) = 7.7 \pm
0.2$ kOe] and $T_{cu} = 9.0 \pm 0.2$ K [$H_{c2}^{(u)}(0) = 6.0 \pm 0.3$
kOe].  The intermediate phase between $T_{cu}$ and $T_{cl}$, where $\delta
\chi (= \chi_{FC} - \chi_{ZFC}) \approx 0$, is an intra-grain
superconductive state occurring in the Ta-C layers in
Ta$_{2}$S$_{2}$C. The low temperature phase below $T_{cl}$, where $\delta
\chi$ clearly appears, is an inter-grain superconductive state.  The
magnetic susceptibility at $H$ well above $H_{c2}^{(l)}(0)$ is described by
a sum of a diamagnetic susceptibility and a Curie-like behavior.  The
latter is due to the localized magnetic moments of conduction electrons
associated with the Anderson localization effect, occurring in the 1T-TaS$_{2}$
type structure in Ta$_{2}$S$_{2}$C.
\end{abstract}


\maketitle



\section{\label{intro}Introduction}
Ta$_{2}$S$_{2}$C has a unique layered structure, where a sandwiched
structure of C-Ta-S-vdw-S-Ta-C is periodically stacked along the $c$ axis. 
\cite{Brec1977,Boller1982,Ziebarath1989,Boller1992,Boller1993,
Wally1998a,Wally1998b,Wally1998c,Walter2000a,Walter2000b}
A van der Waals (vdw) gap is between adjacent S layers.
There are two polytypes: 1T- Ta$_{2}$S$_{2}$C (space
group P\={3}m1, $a = 3.265 \AA$, $c = 8.537 \AA$) and 3R-Ta$_{2}$S$_{2}$C
(space group R\={3}m, $a = 3.276 \AA$, $c = 25.62 \AA$).  In Ta-C-Ta 
layer (for simplicity Ta-C layers are used hereafter),
each [Ta$_{6}$C] octahedron shares six of its 12 edges with adjacent
octahedra and the corner Ta atom is shared by three octahedra.  This edge
linking of [Ta$_{6}$C] octahedra is the common structural feature of
transition metal carbides.  Relatively weak vdW interactions between S
layers give this compound a graphitic character.  The structure of
Ta$_{2}$S$_{2}$C can be viewed as a structural sum of Ta-C layers and TaS$_{2}$,
where the structural part corresponding to Ta-C layers is represented by
[Ta$_{6}$C] octahedra and the structural part corresponding to TaS$_{2}$
is identical to the atom disposition of either 1T-TaS$_{2}$ in the case of
3R-Ta$_{2}$S$_{2}$C or a hypothetical 2H$_{b}$-TaS$_{2}$ (MoS$_{2}$-type)
in the case of 1T- Ta$_{2}$S$_{2}$C. X-ray photoelectron core level
spectra\cite{Walter2000b} of Ta$_{2}$S$_{2}$C show that the binding energy
of carbon is close to that in a graphene sheet on (111) face of Ta rather
than carbon in tantalum carbides (TaC, Ta$_{2}$C).  The binding energy of
sulfur is close to that in 1T-TaS$_{2}$.
The structure of the Ta-C layers in Ta$_{2}$S$_{2}$C is different from 
that of the bulk TaC (cubic) and Ta$_{2}$C (hexagonal) which are cubic 
compounds and do not show any layered structure.   

In this paper we have undertaken an extensive study on the magnetic
properties of Ta$_{2}$S$_{2}$C from SQUID (superconducting interference
device) DC and AC magnetic susceptibility.  We show that this compound
undergoes successive superconducting phase transitions at $T_{cl}$ ($= 3.61
\pm 0.01$ K) and $T_{cu}$ ($= 9.0 \pm 0.2$ K).  The intermediate phase
between $T_{cl}$ and $T_{cu}$ is an intra-grain superconductive state, while
the low temperature phase below $T_{cl}$ is an inter-grain superconductive
state.  The superconducting properties of Ta$_{2}$S$_{2}$C are compared
with those of the type-II superconductor Nb$_{2}$S$_{2}$C with a critical
temperature $T_{c} = 7.6$ K.\cite{Sakamaki2001} In the presence of a
magnetic field ($H$) well above the upper critical field $H_{c2}^{(l)}(0)$ ($=
7.7 \pm 0.2$ kOe), the magnetic susceptibility is described by a sum of a
Curie-like behavior and a diamagnetic susceptibility.  We show that the
Curie-like behavior is due to localized magnetic moments of conduction
electrons associated with the Anderson localization effect, which may occur in
the 1T-TaS$_{2}$ type structure in Ta$_{2}$S$_{2}$C.

The pristine TaC$_{1-x}$ is a superconductor with a fairly high $T_{c}$ (=
9.7 K) for $x \approx 1$, but for $x \approx 0.8$ no superconductivity is
found.\cite{Giorgi1962,Fink1965} The small change of the Fermi surface due
to the nonstoichiometry of the carbides considerably reduces the bulk
phonon anomaly (dip) resulting in the reduction of
$T_{c}$.\cite{Oshima1986} In the pristine 1T-TaS$_{2}$ the charge density
wave (CDW) becomes commensurate with an undistorted host lattice in a first
order transition below 200 K.\cite{Wilson1975} The electrical resistivity
increases about 10-fold at the 200 K transition and below 2 K the
resistivity diverges following the relation $\rho = 
\rho_{0}\exp[(T_{0}/T)^{n}]$
with $n = 1/3$, where $\rho_{0}$ is a constant resistivity and $T_{0}$ 
is a characteristic temperature.\cite{DiSalvo1977} This is characteristic
of the Anderson localization of the conduction electrons due to a random
potential.  The susceptibility shows a Curie-like behavior due to the
localized magnetic moments of conduction
electrons.\cite{DiSalvo1980,Inada1981}

\section{\label{exp}EXPERIMENTAL PROCEDURE}
Powdered samples of Ta$_{2}$S$_{2}$C were prepared by Pablo Wally.  The
detail of the synthesis and structure is described by Wally and 
Ueki.\cite{Wally1998a} X-ray powder diffraction pattern shows that
Ta$_{2}$S$_{2}$C sample consists of a 3R phase as a majority phase and a 1T
phase as a minority phase.\cite{Wally1998a} The sample characterization was
also carried out by scanning tunneling microscopy\cite{Walter2000a} and
X-ray photoelectron spectroscopy.\cite{Walter2000b} The measurements of DC
and AC magnetic susceptibility were carried out using a SQUID magnetometer
(Quantum Design MPMS XL-5).  A polycrystalline powdered sample (mass 253.2
mg) was used in the present work.  Before setting up a sample at 298 K, a
remnant magnetic field was reduced to less than 3 mOe using an ultra-low
field capability option.  For convenience, hereafter this remnant field is
noted as the state $H = 0$.  The detail of the measurements of DC and AC
magnetic susceptibility is described in Sec.~\ref{result}.

\section{\label{result}RESULT}
\subsection{\label{resultA}$\chi^{\prime}$ and $\chi^{\prime\prime}$}

\begin{figure}
\includegraphics[width=8.0cm]{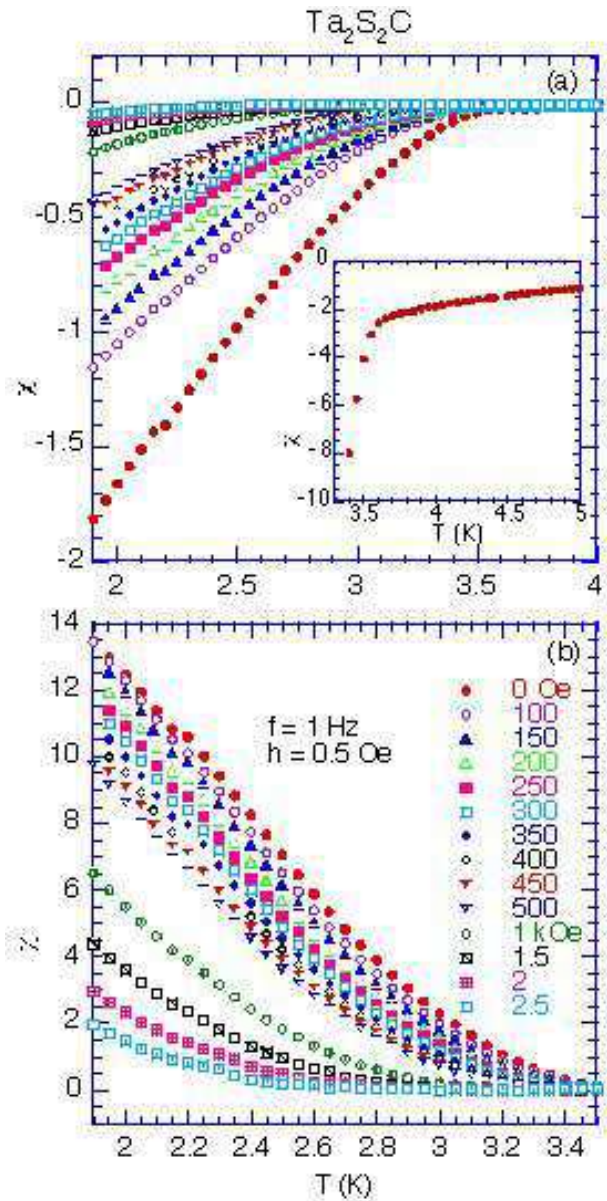}%
\caption{\label{fig:one}$T$ dependence of (a) $\chi^{\prime}$ and (b)
$\chi^{\prime\prime}$ with and without $H$ for Ta$_{2}$S$_{2}$C. $h = 0.5$
Oe.  $f = 1$ Hz.  The inset shows the detail of $\chi^{\prime}$ at $H = 0$
around $T_{cl}$.}
\end{figure}

Figure \ref{fig:one} shows the $T$ dependence of the AC magnetic
susceptibility [(a) the dispersion $\chi^{\prime}$ and (b) the absorption
$\chi^{\prime\prime}$], where the AC frequency $f$ (= 1 Hz) and the
magnitude of the AC field ($h = 0.5$ Oe) are used.  The $T$ dependence of
$\chi^{\prime}$ and $\chi^{\prime\prime}$ is strongly dependent on $H$. 
The sign of $\chi^{\prime}$ is negative at least for $1.9 < T < 5$ K, while
the sign of $\chi^{\prime\prime}$ is positive.  For $H = 0$ (see the inset
of Fig.~\ref{fig:one}(a)), $\chi^{\prime}$ increases with
increasing $T$.  It shows a kink at a critical temperature $T_{cl}$ (= 3.61
K) where the derivative d$\chi^{\prime}$/d$T$ undergoes a discontinuous
jump.  The system undergoes a superconducting transition at $T_{cl}$.  The
dispersion $\chi^{\prime}$ increases with further increasing $T$ and tends
to zero around $T_{cu}$ (= 9.0 K). Similar kink-behavior is observed at
higher $H$, although the kink is hardly seen in Fig.~\ref{fig:one}(a)
because of very small magnitude of $\chi^{\prime}$ at $T_{cl}(H)$.  The
critical temperature $T_{cl}(H)$ decreases with increasing $H$, forming the
$H$-$T$ phase diagram (see Sec.~\ref{resultE}).  The absorption
$\chi^{\prime\prime}$ at $H = 0$ shows a drastic decrease around $T_{cl}$. 
It shows a tail above $T_{cl}$ and is reduced to zero around 5 K. The
temperature at which the tangential line $\chi^{\prime\prime}$ vs $T$ with
the steepest slope intersects the $\chi^{\prime\prime} = 0$ axis coincides
with $T_{cl}$. 

\subsection{\label{resultB}$M$-$H$ loop} 

\begin{figure}
\includegraphics[width=8.0cm]{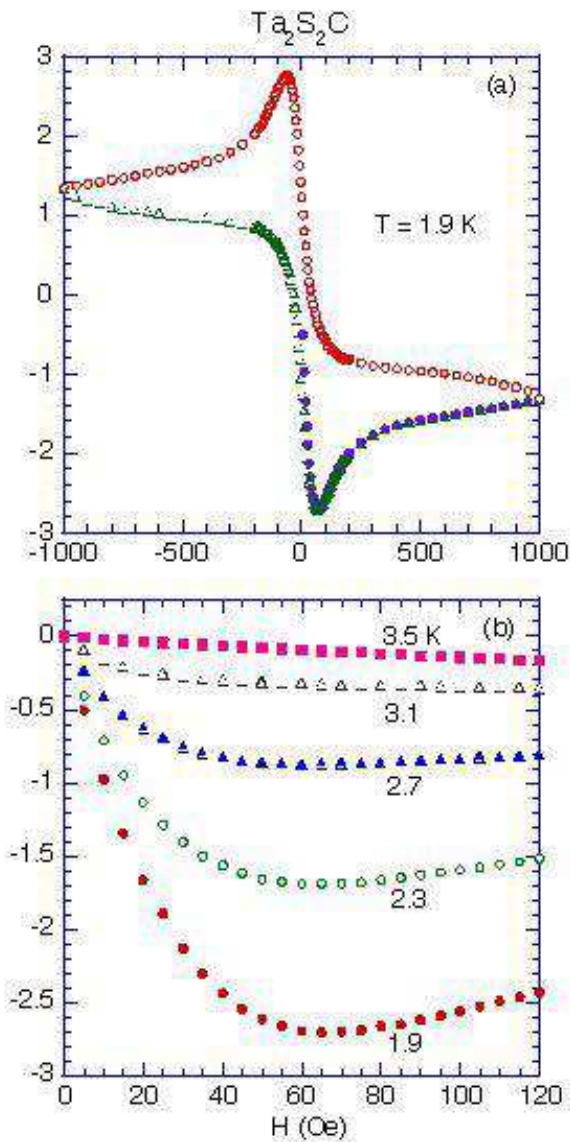}%
\caption{\label{fig:two}(a) $M$-$H$ loop at 1.9 K. (b) $H$ dependence of
$M_{ZFC}$ at various $T$.}
\end{figure}

Figure \ref{fig:two}(a) shows the hysteresis loop of the magnetization $M$
at $T = 1.9$ K. After the sample was quenched from 298 to 1.9 K at $H = 0$,
the measurement was carried out with varying $H$ from 0 to 1 kOe at $T$,
from $H = 1$ to -1 kOe, and from $H = -1$ to 1 kOe.  The $M$-$H$ curve at
1.9 K shows a large hysteresis and a remnant magnetization.  Structural
imperfections or defect in the sample may play an role of flux pinning,
resulting in a inhomogeneous type-II superconductor.  Figure
\ref{fig:two}(b) shows typical data of zero-field cooled (ZFC) 
magmetization $M_{ZFC}$ vs $H$ at various $T$. 
Before each measurement, the sample was kept at 20 K at $H = 0$ for 20
minutes and then it was quenched from 20 K to $T$ ($< 4$ K).  The
magnetization $M_{ZFC}$ at $T$ was measured with increasing $H$ ($0 \leq H
\leq 120$ Oe).  The magnetization $M_{ZFC}$ exhibits a single local minimum
at a characteristic field for $T < T_{cl}$, shifting to the low-$H$ side
with increasing $T$.  The lower critical field $H_{c1}^{(l)}(T)$ is defined
not as the first minimum point of the $M_{ZFC}$ vs $H$, but as the first
deviation point from the linear portion due to the penetration of magnetic
flux into the sample: typically, $H_{c1}^{(l)}(T=1.9$K$) = 20$ Oe and
$H_{c1}^{(l)}(T=2.7$K$) = 12$ Oe.  The least-squares fit of the data of
$H_{c1}^{(l)}$ vs $T$ to a conventional relation\cite{Ketterson1999}
$H_{c1}^{(l)}(T)=H_{c1}^{(l)}(0)[1-(T/T_{cl})^{2}]$ yields $H_{c1}^{(l)}(0)
= 28 \pm 2$ Oe.

\subsection{\label{resultC}$\chi_{ZFC}$ and $\chi_{FC}$}

\begin{figure*}
\includegraphics[width=14.0cm]{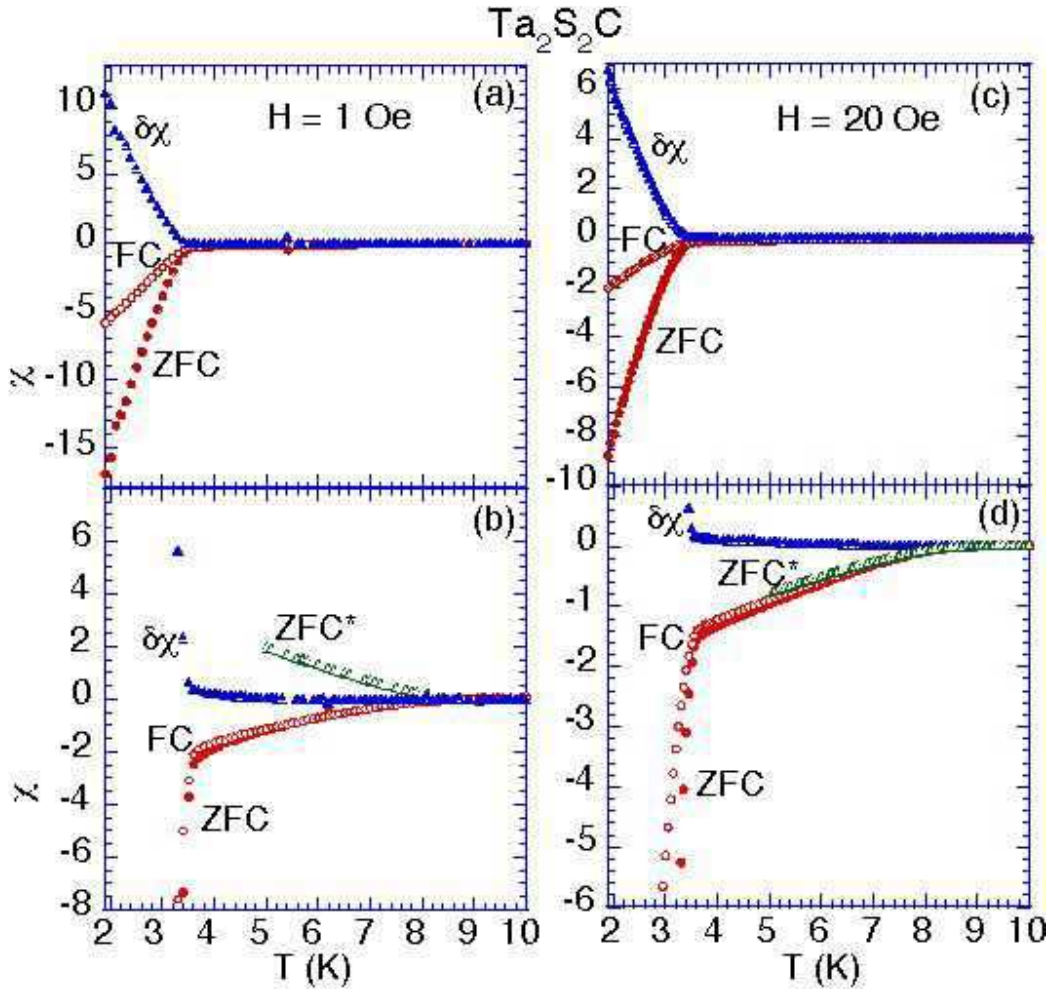}%
\caption{\label{fig:three}$T$ dependence of $\chi_{ZFC}$, $\chi_{FC}$, and
$\delta \chi$ (= $\chi_{FC} - \chi_{ZFC}$).  (a) and (b) for $H = 1$ Oe. 
(c) and (d) for $H = 20$ Oe.  $T$ dependence of $\chi_{ZFC}^{*}$ ($5 \leq T
\leq 10$ K) is also shown for comparison.  The definition of
$\chi_{ZFC}^{*}$ is given in Sec.~\ref{resultD}.}
\end{figure*}

\begin{figure}
\includegraphics[width=8.0cm]{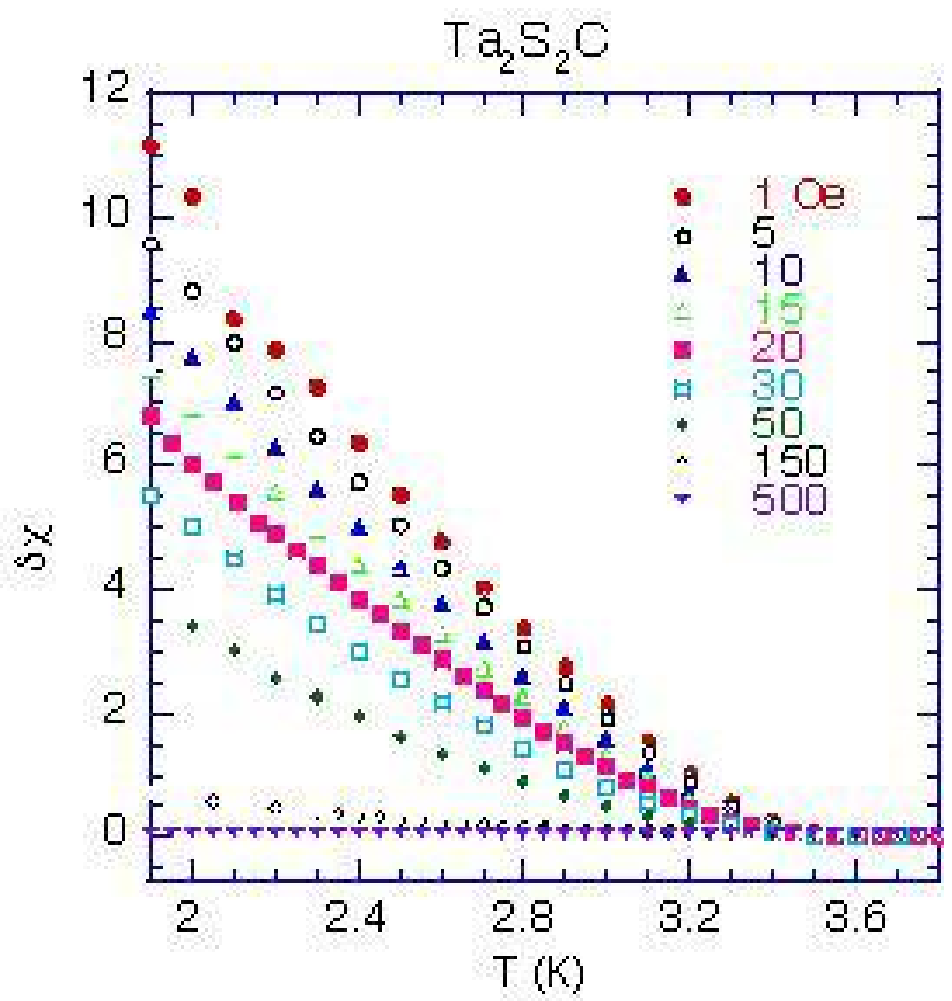}%
\caption{\label{fig:four}$T$ dependence of $\delta \chi$ (=$\chi_{FC} -
\chi_{ZFC}$) at various $H$ around $T_{cl}$.}
\end{figure}

The measurements of $\chi_{ZFC}$ (= $M_{ZFC}/H$) and
$\chi_{FC}$ (= $M_{FC}/H$) were carried out as follows, where $M_{FC}$ 
is the field-cooled (FC) magnetization.  After annealing at
50 K for 1200 sec in the absence of $H$, the sample was quenched from 50
to 1.9 K. The magnetic field $H$ was applied at 1.9 K and then $\chi_{ZFC}$
was measured with increasing $T$.  The sample was again heated up and
annealed at 50 K for 1200 sec in the presence of $H$.  Then $\chi_{FC}$
was measured with decreasing $T$.  Figures \ref{fig:three}(a) and (b)
show the $T$ dependence of $\chi_{ZFC}$, $\chi_{FC}$, and $\delta \chi$ at
$H = 1$ Oe, where $\delta \chi = \chi_{FC} - \chi_{ZFC}$.  The scale of the
susceptibility is enlarged in Fig.~\ref{fig:three}(b).  The susceptibility
$\chi_{ZFC}$ ($\chi_{FC}$) exhibits a kink at $T_{cl}$ (= 3.61 K), where
d$\chi_{ZFC}$/d$T$ (d$\chi_{FC}$/d$T$) undergoes a drastic decrease.  The
deviation of $\chi_{ZFC}$ from $\chi_{FC}$ is clearly seen below $T_{cl}$
(= 3.61 K), indicating that the extra magnetic flux is trapped during the
FC process.  Between $T_{cl}$ and $T_{cu}$ (= 9.0 K), $\delta \chi$ is
still positive but nearly equal to zero.  The sign of $\chi_{ZFC}$
($\chi_{FC}$) changes from negative to positive at $T_{cu}$.  Similar
behavior is also observed at $H$ = 20 Oe as shown in
Figs.~\ref{fig:three}(c) and (d).  Figure \ref{fig:four} shows the $T$
dependence of $\delta \chi$ at various $H$.  The value of $\delta \chi$ at
fixed $T$ decreases with increasing $H$.  The difference $\delta \chi$
undergoes a drastic decrease at $T_{cl}(H)$.  

The minimum value of $\chi_{ZFC}$ at $H = 1$ Oe is $-1.7 \times 10^{-3}$
emu/g at 1.9 K, while the minimum value of $\chi^{\prime}$ at $H = 0$ is
$-1.8 \times 10^{-3}$ emu/g at 1.9 K. Using the value of $\chi^{\prime}$ at
1.9 K ($\approx -1.8 \times 10^{-3}$ emu/g) and the calculated density
$\rho_{cal} = 9.23$ g/cm$^{3}$ for 3R-Ta$_{2}$S$_{2}$C,\cite{Brec1977} the
fraction of flux expulsion relative to complete diamagnetism ($\chi_{0} =
-1/4\pi = -0.0796$ emu/cm$^{3}$) is estimated as 21 \%, suggesting that the
system consists of small grains.  This is in contrast to 38 \% of the
diamagnetic volume fraction reported for Nb$_{2}$S$_{2}$C by
Sakamaki et al.\cite{Sakamaki2001}

\begin{figure}
\includegraphics[width=8.0cm]{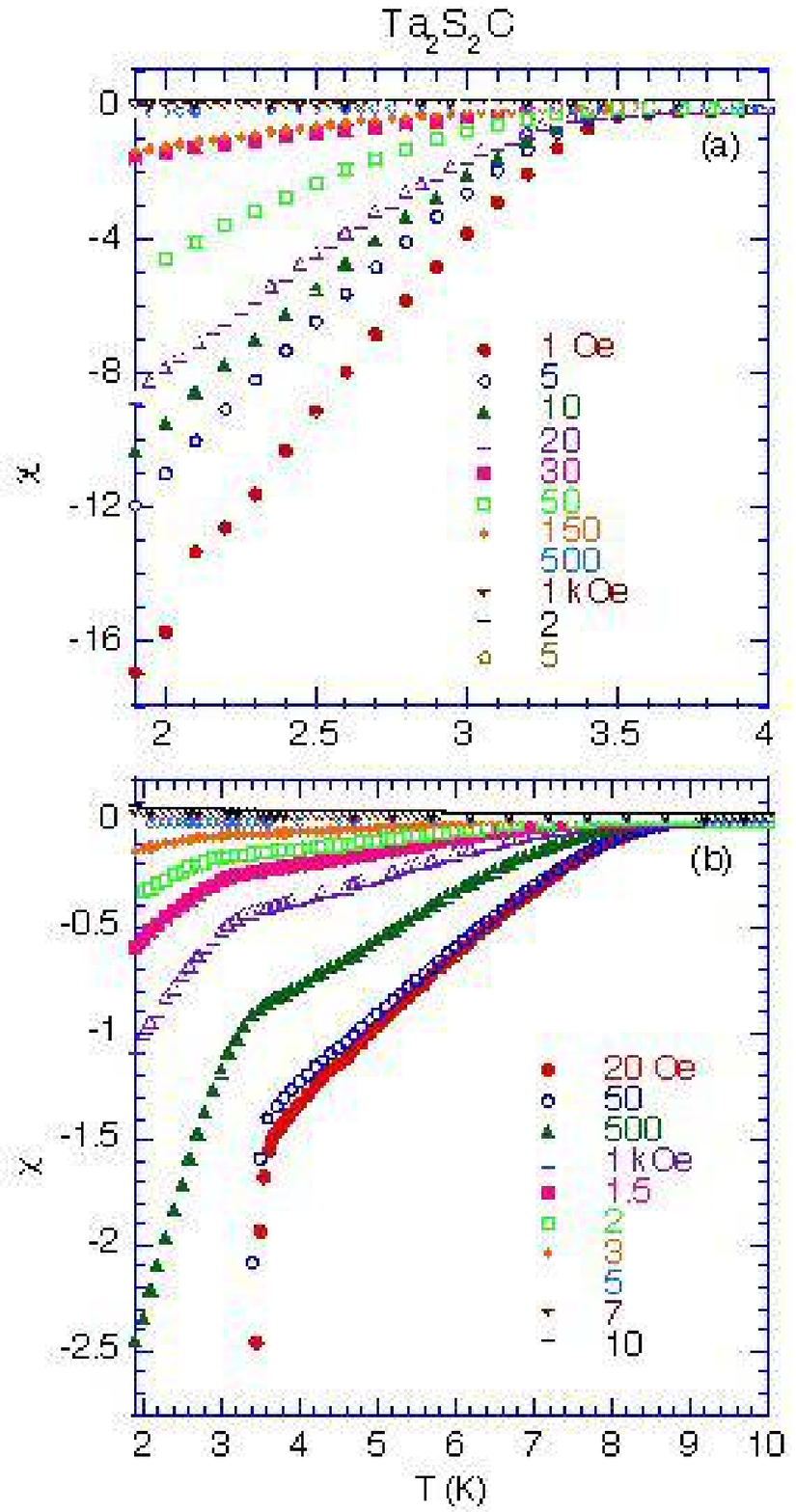}%
\caption{\label{fig:five}(a) and (b) $T$ dependence of $\chi_{ZFC}$ at
various $H$.}
\end{figure}

\begin{figure}
\includegraphics[width=8.0cm]{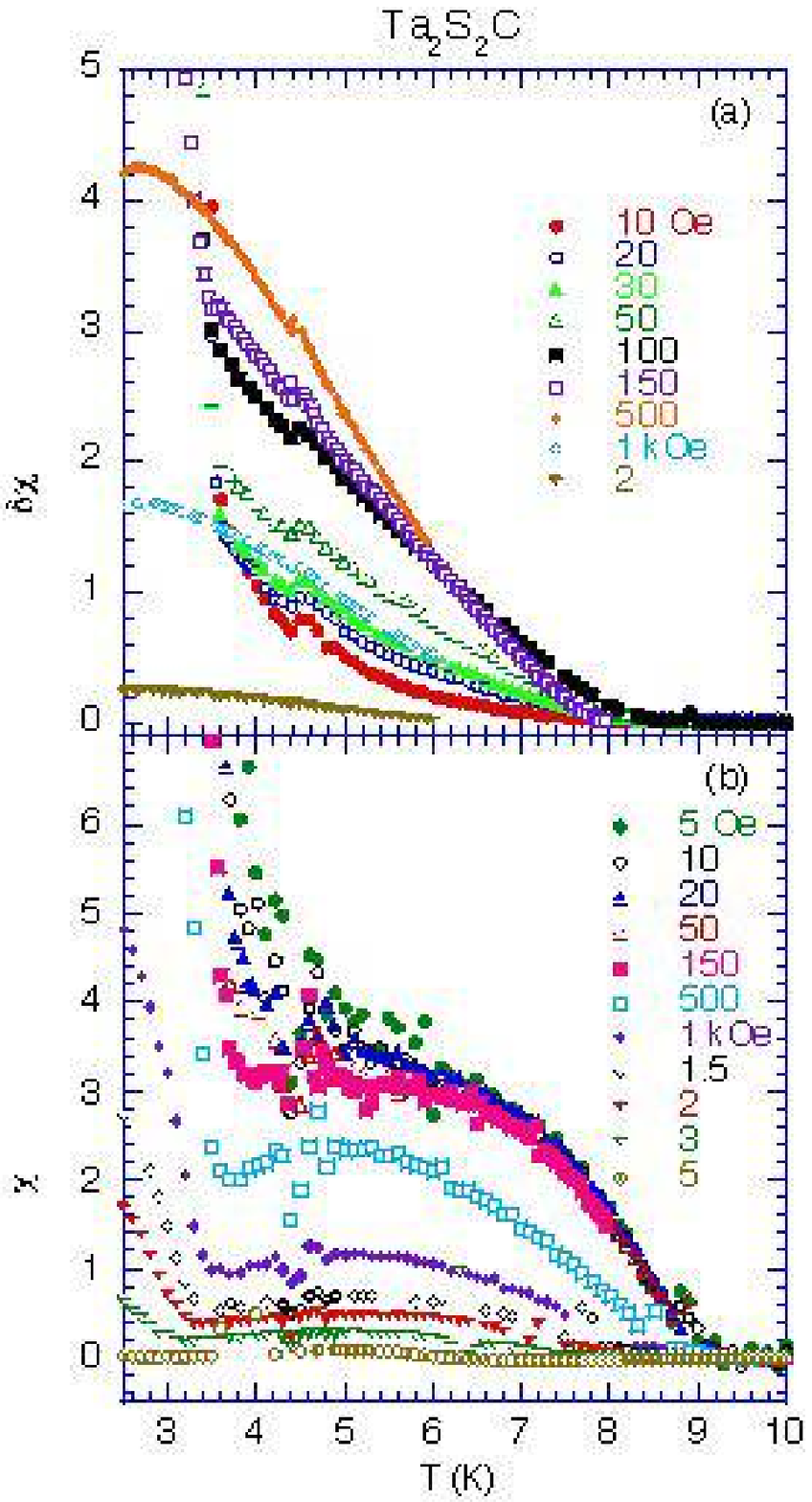}%
\caption{\label{fig:six}$T$ dependence of (a) $\delta \chi$ and (b)
d$\chi_{ZFC}$/d$T$ at various $H$ for $T_{cl} \leq T \leq T_{cu}$.}
\end{figure}

Figures \ref{fig:five}(a) and (b) show the $T$ dependence of $\chi_{ZFC}$ at
various $H$, where the measurement was carried
out between 1.9 and 11 K with increasing $T$.  There is a drastic increase
in the diamagnetic contribution in $\chi_{ZFC}$ associated with the
superconducting transition at $T_{cl}(H)$.  Nevertheless, a diamagnetic
contribution in $\chi_{ZFC}$ still remains above $T_{cl}(H)$, increases with further
increasing $T$, and reduces to a zero at a upper critical temperature
$T_{cu}(H)$.  For $H = 150$ Oe, for example,
$\chi_{ZFC}$ exhibits a kink at $T_{cl}(H)$.  The sign of $\chi_{ZFC}$
changes from negative to positive around 9 K with increasing
$T$.  At $H = 5$ kOe, $\chi_{ZFC}$ is positive at least
between 1.9 and 6 K, showing a broad peak at 2.65 K. At $H = 10$ kOe,
$\chi_{ZFC}$ decreases with increasing $T$, showing a
Curie-like behavior.  Figure \ref{fig:six}(a) shows the $T$ dependence of
$\delta \chi$ at various $H$.  The magnitude of $\delta \chi$ between
$T_{cl}(H)$ and $T_{cu}(H)$ is very small compared to that below 
$T_{cl}(H)$, but
still show the irreversible effect of magnetization.  The difference
$\delta \chi$ at fixed $H$ decreases with increasing $T$ and reduces to
zero at $T_{cu}(H)$.  Note that $\delta \chi$ at fixed $T$ (for example 5
K) increases increasing $H$, showing a maximum around $H = 500$ Oe, and
decreases with further increasing $H$.  This feature is in contrast to the
$H$ dependence of $\delta \chi$ below $T_{cl}(H)$, which decreases with 
increasing $H$. 
Figure \ref{fig:six}(b) shows the $T$ dependence of the derivative
d$\chi_{ZFC}$/d$T$ for 5 Oe $\leq H \leq 10$ kOe.  Clearly
d$\chi_{ZFC}$/d$T$ at low $H$ undergoes two step-like changes around
$T_{cl}(H)$ and $T_{cu}(H)$.  The critical temperatures $T_{cl}(H)$ and
$T_{cu}(H)$ decreases with increasing $H$, forming the $H$-$T$ phase
diagram (see Sec.~\ref{resultE}).

\subsection{\label{resultD}$\chi_{ZFC}^{*}$ in a quasi-equilibrium state}

\begin{figure*}
\includegraphics[width=14.0cm]{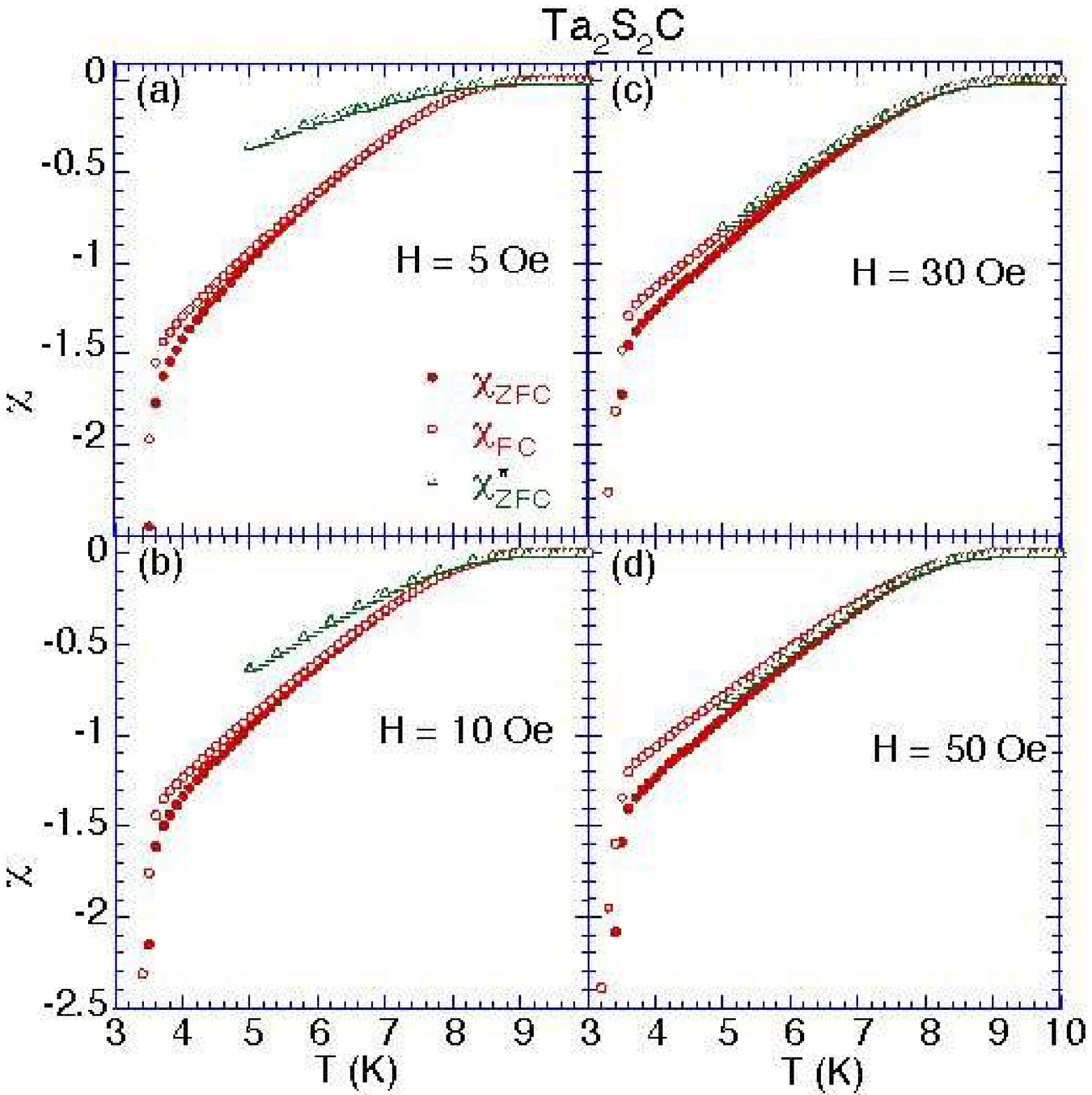}%
\caption{\label{fig:seven}(a)-(d) $T$ dependence of $\chi_{ZFC}^{*}$,
$\chi_{ZFC}$, and $\chi_{FC}$ at various low $H$.  The definition of
$\chi_{ZFC}^{*}$ is given in Sec.~\ref{resultD}.  The measurement of
$\chi_{ZFC}^{*}$ was made with increasing $T$ from $T_{0}$ (= 5 K) after
the sample was quickly heated from 1.9 K (ZFC state) to $T_{0}$.}
\end{figure*}

Here we present our peculiar results on the $T$ dependence of
$\chi_{ZFC}^{*}$.  The method of the measurement for $\chi_{ZFC}^{*}$ is a
little different from that for the conventional $\chi_{ZFC}$.  First, the
sample was annealed at 50 K for 1200 sec in the absence of $H$ and then it
was quenched to 1.9 K. After the sample was kept at 1.9 K for 100 sec in
the presence of fixed $H$, it was quickly heated up to $T_{0}$ between
$T_{cl}$ and $T_{cu}$.  Then $\chi_{ZFC}^{*}$ was measured with increasing
$T$ from $T_{0}$ to 11 K. Figures \ref{fig:seven}(a)-(d) show the $T$
dependence of $\chi_{ZFC}^{*}$, $\chi_{ZFC}$, and $\chi_{FC}$ at various
$H$ ($5 \leq H \leq 100$ Oe), where $T_{0} = 5$ K. The $T$
dependence of $\chi_{ZFC}^{*}$ at the same $H$ is independent of $T_{0}$
when $T_{0}$ is at least between 4 and 8 K. The data of $\chi_{ZFC}^{*}$ vs
$T$ at $H = 1$ and 20 Oe are shown in Figs.~\ref{fig:three}(b) and (d),
respectively.  The susceptibility $\chi_{ZFC}^{*}$  
increases with increasing $T$ and reduces to zero at $T_{cu}(H)$. 
The magnitudes of $\chi_{ZFC}$, $\chi_{ZFC}^{*}$, and $\chi_{FC}$ at the
same $H$ are strongly dependent on $H$: $\chi_{ZFC}^{*} > \chi_{FC} >
\chi_{ZFC}$ for $H = 1 - 30$ Oe, $\chi_{FC} > \chi_{ZFC}^{*} > \chi_{ZFC}$
for $H = 50$ Oe, and $\chi_{FC} > \chi_{ZFC}^{*} = \chi_{ZFC}$ for $H =
100$ and 150 Oe.  Note that $\chi_{ZFC}^{*}$ at $H = 1$ Oe (see
Fig.~\ref{fig:three}(b)), whose sign is positive, decreases with increasing
$T$ and reduces to zero at $T_{cu}$.  Between $T_{cl}$ and $T_{cu}$, the
space of states is divided into at least three states , the ZFC state,
ZFC$^{*}$ state, and FC state.  The system lies in the FC states under the
cooling from 11 K, the ZFC state under the slow heating from 
1.9 K, and the ZFC$^{*}$ state under the rapid heating from 1.9 K.
The value of $H$ (= 30 Oe) is a little higher than the
lower critical field $H_{c1}^{(l)}(T=1.9$K$) = 20$ Oe.  The
susceptibility of each state provides a measure for the corresponding
induced magnetic flux density $B$ which is trapped in the superconducting
grains: $B=H+ 4\pi M$.  The inequality $\chi_{ZFC}^{*} > \chi_{FC}$
indicates that the induced magnetic flux density (or the number of fluxoids
over the system ) in the ZFC$^{*}$ state is higher than that in the FC
state for $H < 30$ Oe.  Such a relatively high flux density in the
ZFC$^{*}$ state may be due to a flux compression as a result of the rapid
redistribution of the grain-pinned vortices which occurs during a change of
$T$ from 1.9 K to $T_{0}$.  This effect exists only at low $H$.  In this
sense, the present effect is similar to the paramagnetic Meissner effect
observed in $\chi_{FC}$ of the pristine Nb\cite{Kostic1996,Pust1998}:
$\chi_{FC}$ at low $H$ becomes positive below the superconducting
transition temperature.  According to Koshelev and
Larkin,\cite{Koshelev1995} the surface supercurrent inhomogeneously trap
the magnetic flux in the sample interior, as a vortex.  In Sec.~\ref{disA}
we discuss a possible distribution of vortices around the grains in the ZFC
and FC states below $T_{cl}$ and between $T_{cl}$ and $T_{cu}$.

\subsection{\label{resultE}$H$-$T$ diagram}

\begin{figure}
\includegraphics[width=8.0cm]{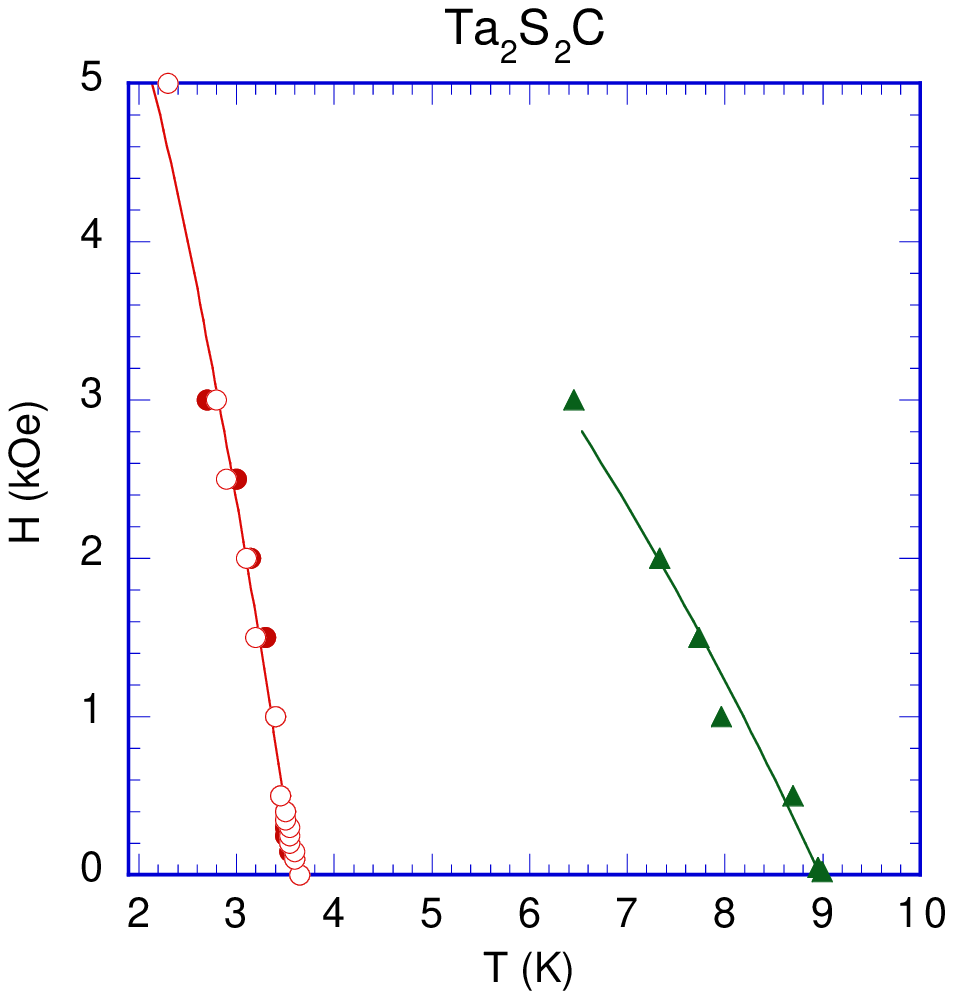}%
\caption{\label{fig:eight}$H$-$T$ phase diagram, where $T_{cl}(H)$ is
determined from the measurements of $\chi^{\prime}$ vs $T$ ({\Large
$\bullet$}) and $\chi^{\prime\prime}$ vs $T$ ({\Large $\circ$}), and
$T_{cu}(H)$ is a temperature where $\chi_{ZFC}$ becomes zero 
($\blacktriangle$).  Solid lines
are least-squares fitting curves for the data of $T_{cl}(H)$ and
$T_{cu}(H)$ to the form $H_{c2}(T) = H_{c2}(0)[1-(T/T_{c})^{2}]$.  The
fitting parameters are given in the text.  $T_{cl} = 3.61$ K. $T_{cu}
= 9.0$ K.}
\end{figure}

Figure \ref{fig:eight} shows the $H$-$T$ phase diagram, where $T_{cl}(H)$
is determined from the data of $\chi^{\prime}$ vs $T$ and
$\chi^{\prime\prime}$ vs $T$, and $T_{cu}(H)$ is determined from the data
of $\chi_{ZFC}$ vs $T$.  These lines correspond to the lines
$H_{c2}^{(l)}(T)$ and $H_{c2}^{(h)}(T)$.  The least squares-fit of the data
of $H$ vs $T$ for the line $H_{c2}^{(l)}(T)$ to a conventional relation
$H_{c2}^{(l)}(T) = H_{c2}^{(l)}(0) [1-(T/T_{cl})^{2}]$ yields
$H_{c2}^{(l)}(0) = (7.7 \pm 0.2)$ kOe and $T_{cl} = 3.61 \pm 0.01$ K, where
the data of $\chi^{\prime}$ vs $T$ and $\chi^{\prime\prime}$ vs $T$ are
used.  The value of $H_{c2}^{(l)}(0)$ thus obtained is comparable to that
estimated using an empirical relation given by Werhammer et
al.,\cite{Werthamer1966} $H_{c2}^{(l)}(T=0$K$)
=-0.69T_{cl}($d$H_{c2}^{(l)}/$d$T)_{T=T_{cl}}$.  In fact, the value of
$H_{c2}^{(l)}(0)$ is calculated as $6.3 \pm 0.2$ kOe, where we use $T_{cl}
= 3.61$ K and a slope (d$H_{c2}^{(l)}/$d$T)_{T_{cl}} = -2500 \pm 189$
(Oe/K) obtained from the linear relation in $H_{c2}^{(l)}$ vs $T$ in the
vicinity of $T = T_{cl}$ and $H_{c2}^{(l)} = 0$.

The coherence length $\xi$ and the magnetic penetration depth $\lambda$ are
related to $H_{c1}^{(l)}(0)$ and $H_{c2}^{(l)}(0)$ through relations
$H_{c1}^{(l)}(0) = \Phi_{0}/(2\pi \xi^{2})$ and $H_{c1}^{(l)}(0) = (\Phi_{0}/4\pi
\lambda^{2}) \ln(\lambda/\xi)$, where $\Phi_{0}$ ($= 2.0678 \times 10^{-7}$
Gauss cm$^{2}$) is the fluxoid.\cite{Ketterson1999} When the values of
$H_{c1}^{(l)}(0)$ (= 27.7 Oe) and $H_{c2}^{(l)}(0)$ (= 7730 Oe) are used,
the values of the Ginzburg-Landau parameter $\kappa$ ($= \lambda/\xi$),
$\lambda$ and $\xi$ can be estimated as $\kappa = 20.5 \pm 0.3$, $\xi = 210
\pm 10 \AA$ and $\lambda = 4200 \pm 100 \AA$.  Our results of $T_{cl}$,
$H_{c1}^{(l)}(0)$, $H_{c2}^{(l)}(0)$, $\kappa$, $\lambda$, and $\xi$ in
Ta$_{2}$S$_{2}$C thus obtained are compared to those in
Nb$_{2}$S$_{2}$C\cite{Sakamaki2001}: $T_{c} = 7.6$ K, $H_{c1}(0) = 227 \pm
4$ Oe and $H_{c2}(0) = 9950 \pm 180$ Oe, $\kappa = 6.37$, $\xi = 182 \pm 2
\AA$, and $\lambda = 1160 \pm 20 \AA$.  Both the pristine Nb and Ta are
superconductive elements with the critical temperatures $T_{c}($Nb$) = 9.25
\pm 0.02$ K and $T_{c}($Ta$) = 4.47 \pm 0.04$ K,\cite{Ketterson1999} and
have a body-centered cubic structure.  We find that the ratio
$T_{c}$(Nb$_{2}$S$_{2}$C)/$T_{c}$(Ta$_{2}$S$_{2}$C) (= 2.11) is close 
to the ratio $T_{c}$(Nb)/$T_{c}$(Ta) (= 2.06).  The values of $T_{c}$
and $H_{c2}$ for Ta$_{2}$S$_{2}$C are comparable to those of
2$H_{a}$-TaS$_{x}$Se$_{2-x}$ ($0.4 < x < 1.8$): $T_{c} = 3.9$ K and $H_{c2}
= 6.7$ kOe for $x = 0.8$, $T_{c} = 3.7$ K and $H_{c2} = 9 - 11$ kOe for $x
= 1$, $T_{c} = 3.9$ K and $H_{c2} = 11.5 - 12.8$ kOe at $x =
1.2$.\cite{Morris1973}

What kind of the superconductivity occurs at $T_{cu}(H)$?  The least
squares-fit of the data of $H$ vs $T$ for the line $H_{c2}^{(u)}(T)$ to a
conventional relation $H_{c2}^{(u)}(T) = H_{c2}^{(u)}(0)
[1-(T/T_{cu})^{2}]$ yields $H_{c2}^{(u)}(0) = (6.0 \pm 0.3)$ kOe and
$T_{cl} = 8.98 \pm 0.06$ K, where the data of $\chi_{ZFC}$ vs $T$ are used. 
The origin of the superconductivity at $T_{cu}(H)$ may be due to Ta-C layers
in Ta$_{2}$S$_{2}$C. According to Giorgi et al.,\cite{Giorgi1962}
the critical temperature $T_{c}$ of the pristine TaC$_{1-x}$ increases with
decreasing $x$ and is equal to 9.0 K at $x = 0.019$.  Fink et
al.\cite{Fink1965} have reported that the values of $H_{c1}$ and $H_{c2}$
at $T = 1.2$ K for TaC are 220 Oe and 4.6 kOe.  These values are in good
agreement with the values of $T_{cu}$ and $H_{c2}^{(u)}(0)$ in
Ta$_{2}$S$_{2}$C. The origin of the superconductivity at $T_{cl}$ and
$T_{cu}$ will be discussed in Sec.~\ref{disA}.

\subsection{\label{resultF}$\chi_{ZFC}$ and $\chi_{FC}$ at high $H$ 
and high $T$}

\begin{figure}
\includegraphics[width=8.0cm]{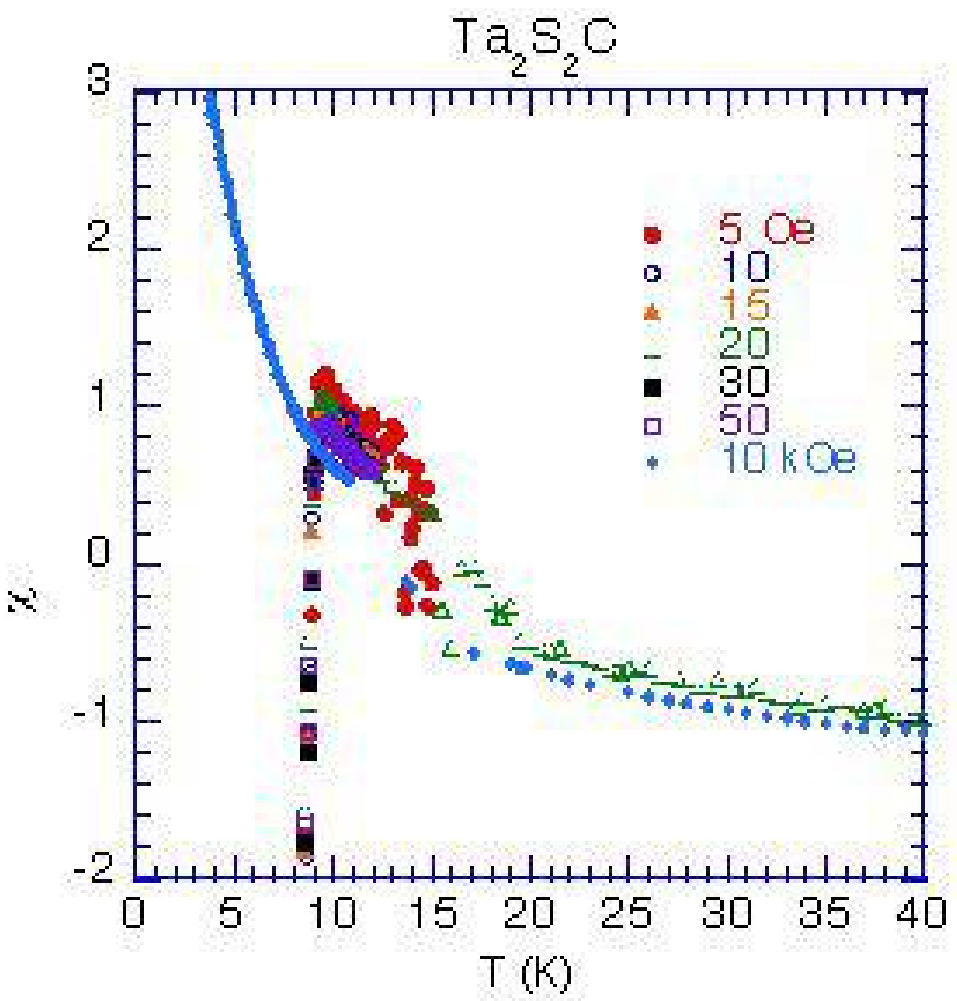}%
\caption{\label{fig:nine}$T$ dependence of $\chi_{FC}$ at various $H$ near
$T_{cu}$.}
\end{figure}

In Fig.~\ref{fig:nine} we show the $T$ dependence of $\chi_{ZFC}$ and
$\chi_{FC}$ at various $H$.  The susceptibility $\chi_{FC}$ (also
$\chi_{ZFC}$) at $H = 20$ Oe show a sharp peak around 9.5 K and decreases
with further increasing $T$. The deviation of $\chi_{FC}$ from
$\chi_{ZFC}$ is observed for $15 \leq T \leq 19$ K. The sign of $\chi_{FC}$
changes from positive to negative around 16 K. The discontinuous jump of
$\chi$ observed in the vicinity of $\chi = 0$ is an artifact due to the
SQUID measurement.  In contrast, $\chi_{FC}$ (= $\chi_{ZFC}$) at $H = 10$
kOe decreases with increasing $T$ from 1.9 K and merge to the curves of
$\chi_{FC}$ at $H = 20$ Oe above 9.5 K.

\begin{figure}
\includegraphics[width=8.0cm]{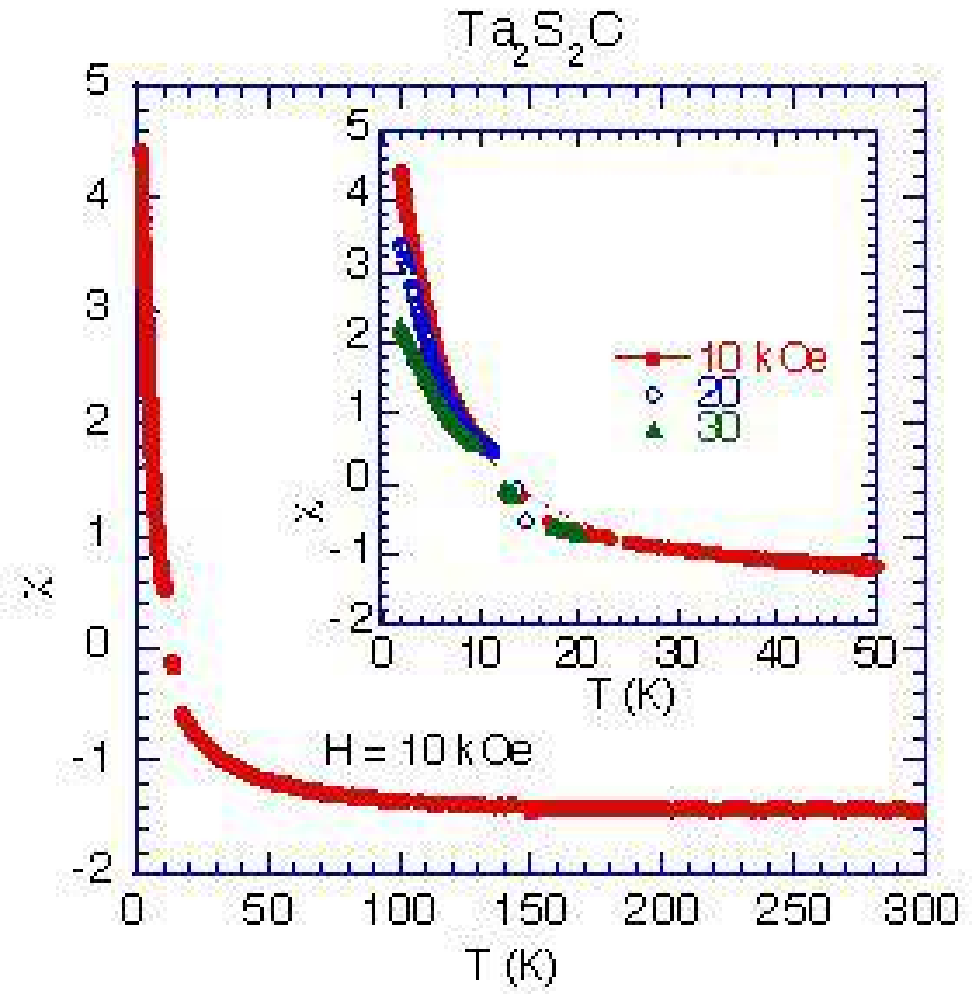}%
\caption{\label{fig:ten}$T$ dependence of $\chi_{FC}$ at $H = 10$ kOe.  The
inset shows the $T$ dependence of $\chi_{FC}$ at $H = 10$, 20, and 30 kOe. 
The solid line in the inset denotes the least-squares fitting curve to
Eq.(\ref{eq:one}).  The fitting parameters are given in the text.}
\end{figure}

Figure \ref{fig:ten} shows the $T$ dependence of $\chi_{FC}$ at $H = 10$
kOe for $1.9 \leq T \leq 298$ K. The susceptibility $\chi$ consists of a
diamagnetic susceptibility and a Curie-like susceptibility.  The
susceptibility $\chi_{FC}$ slightly decreases with increasing $T$ above 20
K and reaches a negative constant (diamagnetic susceptibility) above 150 K:
$\chi_{d} = (-1.44 \pm 0.01) \times 10^{-7}$ emu/g at 298 K. Similar $T$
dependence of $\chi$ is observed in 1T-TaS$_{2}$ below 150
K.\cite{DiSalvo1980,Inada1981} A step-like change $\chi$ near 200 K for
1T-TaS$_{2}$ is due to the CDW transition between a high-temperature
incommensurate phase and a low-temperature commensurate phase.  As seen in
Fig.~\ref{fig:ten} there is no such anomaly above 20 K in Ta$_{2}$S$_{2}$C.
The inset of Fig.~\ref{fig:ten} shows the $T$ dependence of $\chi_{FC}$ at
$H = 10$, 20, and 30 kOe.  The susceptibility is strongly dependent on $H$
below 10 K, indicating that $M_{FC}$ is nonlinear in $H$ (see also the data
of $M_{ZFC}$ vs $H$ in Fig.~\ref{fig:eleven}).  Note that the discontinuous
jump of $\chi_{FC}$ observed in the vicinity of $\chi_{FC} = 0$ is an
artifact due to the SQUID measurement.  Almost all the data between 10 and
16 K are removed from the figure.  The least-squares fit of the data of
$\chi_{FC}$ vs $T$ at $H = 10$ kOe for $4 \leq T \leq 43$ K to
\begin{equation}
\chi_{FC}=\chi_{0} + C/(T - \Theta),
\label{eq:one}
\end{equation}
yields the $T$-independent susceptibility $\chi_{0} = (-1.80 \pm 0.02)
\times 10^{-7}$ emu/g, the Curie-Weiss constant $C = (2.91 \pm 0.04) \times
10^{-6}$ emu K/g, and the Curie-Weiss temperature $\Theta$ = -2.37 $\pm$
0.07 K. The magnitude of $\chi_{0}$ is a little larger than that of
$\chi_{d}$ determined above.  The susceptibility $\chi_{FC}$ consists of a
Curie-like behavior at low $T$ and a diamagnetic contribution at high $T$. 
We assume that the Curie-like behavior is due to the localized electron
spins having the effective magnetic moment $P_{eff} =
g[S(S+1)]^{1/2}\mu_{B} = \sqrt{3} \mu_{B}$, where the Land\'{e} g-factor
$g$ = 2 and the spin $S = 1/2$.  This localized magnetic moment could be
related to the Anderson localization (see Sec.~\ref{disB}).  Through the
repulsive Coulomb interaction between the electrons, a singly occupied
state exists in the vicinity of $\epsilon_{F}$.  Then the number of spins
per gram of Ta$_{2}$S$_{2}$C ($N_{g}$) can be estimated as $N_{g} = 4.67
\times 10^{18}$/g (or $4.3 \times 10^{9}$/cm$^{3}$) from the Curie-Weiss
constant.  In summary, the superconductivity below $T_{cl}$, which is
dominant at low $H$, is weakened as $H$ increases and overcome by the
localization effect at high $H$ well above $H_{c2}^{(l)}(0)$.

There may be another possibility that the Curie-Weiss behavior is due to
magnetic impurities (for example, Fe$^{2+}$), which may be contained in
original Ta (typically 15 ppm Fe in the pristine
1T-TaS$_{2}$).\cite{Inada1981} However, this possibility may be ruled out
in the following way.  If each Fe$^{2+}$ ion has the effective magnetic
moment ($= 5.4 \mu_{B}$), then the number of Fe$^{2+}$ spins per gram of
Ta$_{2}$S$_{2}$C is estimated as $N_{g} = 2.0 \times 10^{19}$ /g.  The
magnitude of Fe$^{2+}$ impurities is estimated as 1800 ppm, which is too
large compared to $\approx 15$ ppm as major metallic impurities of
Ta$_{2}$S$_{2}$C. Similar Curie-Weiss behavior is observed in the
susceptibility of 1T-TaS$_{2}$: $\chi_{0} = -1.96 \times 10^{-7}$ emu/g,
$\Theta = 0.4$ K, and $C = 0.806 \times 10^{-6}$ emu K/g.\cite{DiSalvo1980}
These values of $\chi_{0}$, $\Theta$, and $C$ are comparable to those
derived in the present work for Ta$_{2}$S$_{2}$C. The Curie-like behavior
in 1T-TaS$_{2}$ is not due to magnetic impurities, but due to the localized
magnetic moments related to the Anderson localization
effect.\cite{DiSalvo1980,Inada1981} The diamagnetic susceptibility is
common to the 1T-polytypes showing CDW's.  In 1T-TaS$_{2}$, the effect of
the ordering of the commensurate CDW below 200 K is clearly observed in a
discontinuous jump in the electrical resistivity.\cite{DiSalvo1980,Inada1981} In
Ta$_{2}$S$_{2}$C the $T$ dependence of the electrical resistivity $\rho$
has been reported by Ziebarath et al.\cite{Ziebarath1989} using a
pressed- and sintered-sample.  The resistivity increases with increasing $T$
for $4.2 \leq T \leq 300$ K ($\rho \approx 0.8$ m$\Omega$cm at 4.2 K and
2.2 m$\Omega$cm at 300 K), showing a metallic behavior.  Unlike the
resistivity of 1T-TaS$_{2}$, no discontinuous change in $\rho$ has been
observed below 300 K.

\subsection{\label{resultG}$M_{ZFC}$ vs $H$ at low $T$ and high $H$}

\begin{figure}
\includegraphics[width=8.0cm]{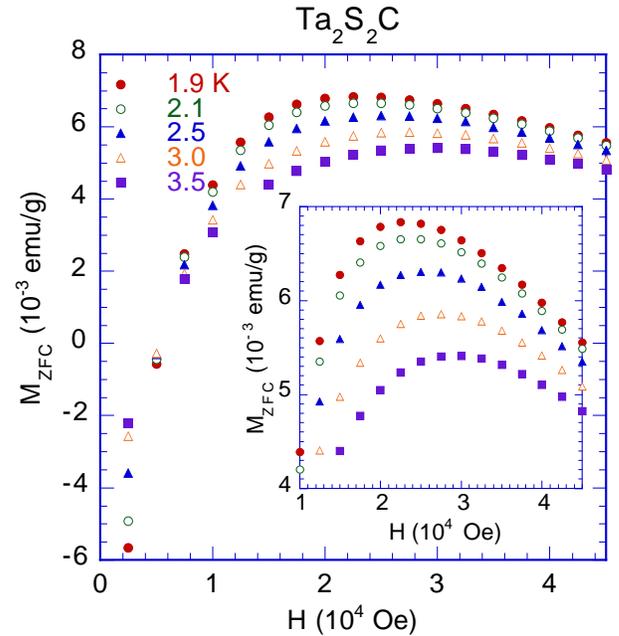}%
\caption{\label{fig:eleven}Magnetization $M_{ZFC}$ as a function of $H$ at
low $T$.}
\end{figure}

\begin{figure}
\includegraphics[width=8.0cm]{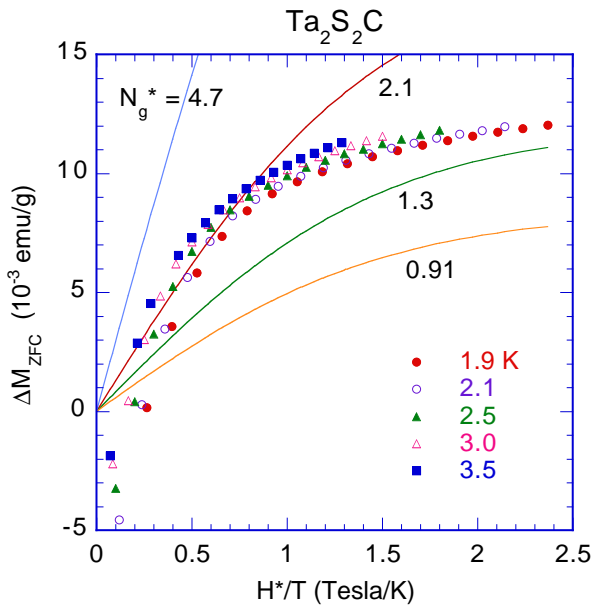}%
\caption{\label{fig:twelve}Difference $\Delta M_{ZFC}$ ($= M_{ZFC} - \chi_{d}H$) as a
function of $H^{*}/T$ where $\chi_{d} = -1.44 \times 10^{-7}$ emu/g, $H =
H^{*} \times 10^{4}$ ($H$ and $H^{*}$ are in the units of Oe and Tesla,
respectively), and the data of Fig.~\ref{fig:eleven} for $M_{ZFC}$ are
used.  The solid lines denote Brillouin functions given by
$N_{g}\mu_{B}\tanh(0.6717 H^{*}/T)$ with $N_{g} = (4.7, 2.1, 1.3, 0.91)
\times 10^{18}$/g.}
\end{figure}

Figure \ref{fig:eleven} shows the $H$ dependence of $M_{ZFC}$ at fixed $T$
below $T_{cl}$.  The susceptibility $M_{ZFC}$ is negative at low $H$
because of the Meissner effect.  The sign of $M_{ZFC}$ changes to positive
at a zero-crossing field $H_{0}(T)$.  The value of $H_{0}$ ($= 5.6$ kOe 
at 1.9 K)
coincides with that evaluated from the relation $H_{c2}^{(l)}(T) =
H_{c2}^{(l)}(0) [1-(T/T_{cl})^{2}]$.  The magnetization $M_{ZFC}$ increases
with further increasing $H$.  It shows a broad peak, which shifts to the
high-$H$ side with increasing $T$: 22.5 kOe at 1.9 K and 29 kOe at 3.5 K. 
This peak arises from a competition between the Curie-like behavior
($\Delta M_{ZFC}$) and the diamagnetic contribution ($\chi_{d}H$).  In
Fig.~\ref{fig:twelve} we show the plot of a magnetization $\Delta M_{ZFC}$
as a function of $H^{*}/T$, where $\Delta M_{ZFC} = M_{ZFC} - \chi_{d}H$
with $\chi_{d} = -1.44 \times 10^{-7}$ emu/g, and $H = 10^{4}H^{*}$ ($H$ and $H^{*}$ are
in the units of Oe and Tesla, respectively).  The curve of $\Delta M_{ZFC}$
vs $H^{*}/T$, which are slightly different for different $T$, increases
with increasing $H^{*}/T$.  The function form of $\Delta M_{ZFC}$ vs
$H^{*}/T$ will be discussed in Sec.~\ref{disB}.

\section{\label{dis}DISCUSSION}
\subsection{\label{disA}Origin of successive phase transitions at $T_{cl}$
and $T_{cu}$}
We find that Ta$_{2}$S$_{2}$C undergoes successive superconducting
transitions at $T_{cl}$ and $T_{cu}$.  The diamagnetic volume fraction (21
\%) indicates that our system is formed of many small grains.  The $T$ and
$H$ dependence of $\chi_{ZFC}$ and $\chi_{FC}$ below $T_{cl}$ is very
different from that between $T_{cl}$ and $T_{cu}$: $\delta \chi > 0$ for $T
< T_{cl}$ and $\delta \chi \approx 0$ ($\delta \chi > 0$) for $T_{cl} < T <
T_{cu}$.  The difference $\delta \chi$ at a fixed $T$ below $T_{cl}$
decreases with increasing $H$.  In contrast, $\delta \chi$ at a fixed $T$
between $T_{cl}$ and $T_{cu}$ increases with increasing $H$, showing a
maximum around $H = 500$ Oe, and decreases with further increasing $H$.  We
note that similar successive transitions have been reported in a ceramic
superconductor YBa$_{2}$Cu$_{4}$O$_{8}$ ($T_{cl} = 37$ K and $T_{cu} = 80$
K),\cite{Kawachi1994,Matsuura1995} which consists of small grains.  Below
$T_{cu}$, $\chi_{ZFC}$ ($= \chi_{FC}$) becomes negative.  The difference
$\delta \chi$ appears below $T_{cl}$.  The difference $\delta \chi$ at a
fixed $T$ below $T_{cl}$ decreases with increasing $H$.

The successive phase transitions at $T_{cl}$ and $T_{cu}$ in
YBa$_{2}$Cu$_{4}$O$_{8}$ can be qualitatively explained by Kawachi et
al.\cite{Kawachi1994} by taking into account of the role of the
superconducting grain below $T_{cu}$ and clusters below $T_{cl}$. 
According to their model, our results of Ta$_{2}$S$_{2}$C can be explained
as follows.  Below $T_{cu}$ the superconductivity occurs in each grain.  In
the presence of $H$ well above $H_{c1}^{(u)}$, the fluxoids are pinned by
pinning centers such as defects and vacancies within grains.  A very weak
irreversible effect of magnetization suggests that the density of fluxoids
pinned in each grain for the FC state is slightly larger than the ZFC
state.  The distribution of fluxoids in the FC state is more uniform, while
in the ZFC state more fluxoids are concentrated in the grain-boundary,
which does not contributes to
$\chi_{ZFC}$.\cite{Rzchowski1990a,Rzchowski1990b} Below $T_{cl}$, the
superconductivity occurs in clusters formed of grains coupled through weak
inter-grain Josephson couplings.  In the presence of $H$, the system is
divided into relatively free regions (the cluster-boundary region) and
strong-pinned regions (within the clusters).  In the FC state, the
distribution of the fluxoids is more uniform inside the clusters.  In the
ZFC state, more fluxoids are concentrated in the cluster-boundary regions
which may have little contribution to $\chi_{ZFC}$.  In the presence of $H$
($< H_{c1}^{(l)}$) applied to the system in the ZFC state, the fluxoids
will enter only the cluster-boundary regions around the clusters.  The
fluxoids in the inter-cluster region can have a path through the system
without being caught by clusters.  When $H > H_{c1}^{(l)}$, some free
fluxoids are caught by the clusters and pinned strongly, which contributes
to $\chi_{ZFC}$.  Because of such an increase in $\chi_{ZFC}$, $\delta
\chi$ decreases with increasing $H$.

The possible existence of mesoscopic grains in the Ta-C layers of
Ta$_{2}$S$_{2}$C would be essential to the successive transitions having a
hierarchical nature.  The superconductive ordering proceeds in two steps
from the intraplanar Josephson couplings between grains in the same Ta-C
layers to the interplanar Josephson interaction between grains in adjacent
Ta-C layers separated by TaS$_{2}$-type structure.  In the intermediate
phase between $T_{cu}$ and $T_{cl}$, each grain in Ta-C layers becomes a
superconductor.  Through the intraplanar Josephson coupling between grains,
the region of the superconducting grains becomes larger as $T$ decreases
below $T_{cu}$, forming a 2D superconducting phase.  Thermal fluctuations
overcome the interplanar Josephson coupling.  Just below T$_{cl}$, the
effective interplanar Josephson coupling becomes strong enough to give rise
to a 3D superconducting phase.  The 2D superconducting systems are coupled
to each other through a weak interplanar Josephson coupling between
adjacent Ta-C layers.

We note that the magnetic analogy to the successive phase transitions of
such a hierarchical nature is seen in a stage-2 CoCl$_{2}$ graphite
intercalation compound (GIC):\cite{Suzuki2002} $T_{cu}$ (= 8.9 K) and
$T_{cl}$ ($= 6.8 - 7.2$ K).  The existence of islands is essential to the
successive phase transitions.  The nearest neighbor spins inside islands
are ferromagnetically coupled with intraplanar exchange interactions.  On
approaching $T_{cu}$ from the high-$T$ side, spins come to
order ferromagnetically inside islands.  At $T_{cu}$ these ferromagnetic
islands continue to order over the same layer through interisland
interactions (mainly ferromagnetic), forming a 2D ferromagnetic long-range
order.  Below $T_{cl}$ a reentrant spin-glass-like phase order is
established through effective antiferromagnetic interplanar interactions
between spins in adjacent intercalate layers, where the antiferromagnetic
phase and the spin glass phase coexist.

\subsection{\label{disB}Anderson localization effect}

\begin{figure}
\includegraphics[width=8.0cm]{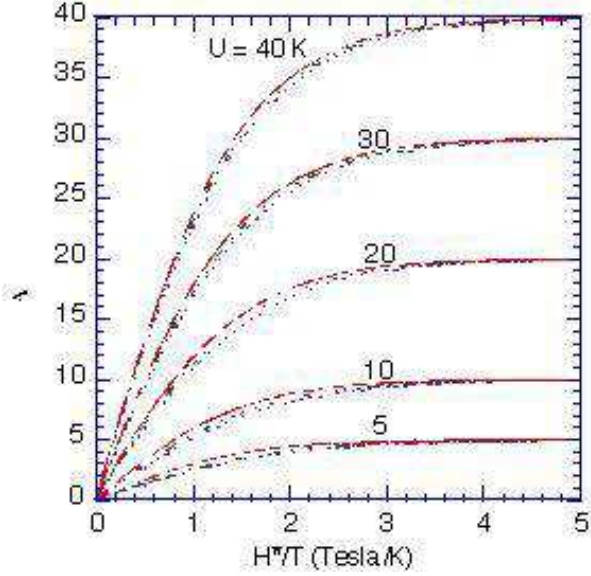}%
\caption{\label{fig:thirteen}Plot of $\Delta M^{*} = (M_{e} -
M_{P})/(k_{B}N(0)\mu_{B})$ for $T = 1.9$ K (dash-dotted line) and $T = 3.5$ K
(dotted line), and the Brillouin function given by $M_{B}^{*} =
U\tanh(\mu_{B}H/k_{B}T)$ (solid line) as a function of $H^{*}/T$ as the
intrastate Coulomb interaction $U$ (in the units of K) is changed as a
parameter.  $U = 27$ K for 1T-TaS$_{2}$.\cite{Inada1981} $M_{e}$ is defined
by Eq.(\ref{eq:four}).  $M_{P}$ is the Pauli paramagnetism given by $M_{P}
= 2N(0)\mu_{B}^{2}H$.}
\end{figure}

We have shown that in Ta$_{2}$S$_{2}$C, the susceptibility $\chi_{FC}$ at
low $T$ and high $H$ obeys the Curie law, which is a direct evidence of the
appearance of local magnetic moments of unpaired electrons due to the 
Anderson localization effect.  A theory
for the magnetic susceptibility in systems with both localization and
electron-electron interactions is presented by Kobayashi et
al.,\cite{Kobayashi1978} and Kamimura and Aoki.\cite{Kamimura1989} It is
assumed that there are localized states with energies very close to the
Fermi level $\epsilon_{F}$.  Singly occupied states are energetically
favorable by virtue of intra-state Coulomb interaction ($U > 0$) between
spin-up and spin-down electrons in the same localized state.  In the
presence of $H$, the magnetization contribution $m_{i}$ of the $i$-th
localized state with energy $\epsilon_{i}$, relative to $\epsilon_{F}$, can
be written as
\begin{equation}
m_{i} = \mu_{B} [e^{-\beta(\epsilon_{i}-\mu_{B}H)}- 
e^{-\beta(\epsilon_{i}+\mu_{B}H)}]/Z
\label{eq:two}
\end{equation}
where $\beta = 1/(k_{B}T)$, $\mu_{B}$ is the Bohr magneton, and $Z$ is the
partition function given by
\begin{equation}
Z=1+e^{-\beta(\epsilon_{i}-\mu_{B}H)}+e^{-\beta(\epsilon_{i}+\mu_{B}H)} 
+ e^{-\beta(2\epsilon_{i}+U)},
\label{eq:three}
\end{equation}
corresponding to an empty state, a state occupied by a spin-up electron
with the energy $\epsilon_{i}$ - $\mu_{B}H$, a state occupied by a
spin-down electron with the energy $\epsilon_{i}$ + $\mu_{B}H$, and a state
occupied by spin-up and spin-down electrons with 2 $\epsilon_{i}$ +$U$. 
Then the total magnetization $M_{e}$ is obtained as
\begin{widetext}
\begin{equation}
M_{e} =\frac{N(0)\mu_{B} \sinh (\beta \mu_{B} H)} {\beta \lbrack \cosh^{2}
(\beta \mu _{B} H)-e^{-\beta U} ]^{1/2} }
\ln {\frac{\cosh (\beta \mu_{B}
H)+\lbrack \cosh ^{2} (\beta \mu _{B} H)-e^{-\beta U} ]^{1/2}}{\cosh
(\beta \mu _{B} H)-\lbrack \cosh ^{2} (\beta \mu _{B} H)-e^{-\beta U}
]^{1/2} } },
\label{eq:four}
\end{equation}
\end{widetext}
where $N(0)$ is the density of states at $\epsilon_{F}$.  For $U = 0$, $M$
is equal to the Pauli paramagnetism $M_{P} = 2N(0)\mu_{B}^{2}H$.  Here we
define $\Delta M^{*}$ as $\Delta M^{*} = (M_{e} -
M_{P})/(k_{B}N(0)\mu_{B})$, where $H$ is in the unit of Oe and $U$ is in the
unit of K: $U = 27$ K for the pristine 1T TaS$_{2}$.  Figure
\ref{fig:thirteen} shows the plot of $\Delta M^{*}$ as a function of
$H^{*}/T$ for $0 \leq H^{*} \leq 10$ Tesla and $T = 1.9 $ and $3.5$ K as $U$ is
changed as a parameter, where $H^{*}$ is in the units of Tesla.  For
comparison, the Brillouin function given by $M_{B}^{*} = U \tanh(\beta
\mu_{B}H)$ is also plotted for each $U$.  Although $\Delta M^{*}$ depends
on both $H^{*}/T$ and $U/T$ for each $U$, the curve of $\Delta M^{*}$ vs
$H^{*}/T$ for each $U$ fits well with $M_{B}^{*}$ with the same $U$.  Note
that $M_{B}^{*}$ is a little larger than $\Delta M^{*}$ at the same
$H^{*}/T$ for 0$ \leq H^{*}/T \leq $3.  The magnetization $\Delta M^{*}$
at $H^{*}/T = 5$ is equal to the saturation magnetization ($= U$) of
$M_{B}^{*}$.  This implies that $\Delta M^{*}$ with $U$ coincides with the
magnetization of free electron spins whose number is given by $N(0)U$.

Here we discuss our result shown in Fig.~\ref{fig:twelve} based on the
above model.  As shown Fig.~\ref{fig:twelve}, all the data of $\Delta
M_{ZFC}$ vs $H^{*}/T$ do not fall on a single-valued function of $H^{*}/T$.
This is consistent with the expression given by Eq.(\ref{eq:four}). 
Negative values of $\Delta M_{ZFC}$ for $H^{*}/T < 0.3$ is due to the
Meissner effect.  In Fig.~\ref{fig:twelve} a solid line denotes a
Brillouin function given by $N_{g}\mu_{B} \tanh(0.6717 H^{*}/T)$ with
$N_{g}\mu_{B} = 0.019$ emu/g or $N_{g} = 2.1 \times 10^{18}$/g, where
$N_{g} = N(0)U$.  Our data of $\Delta M_{ZFC}$ vs $H^{*}/T$ greatly
deviates downward from this Brillouin function for $H^{*}/T > 1$.  
In Fig.~\ref{fig:twelve} the
magnetization $\Delta M_{ZFC}$ reaches 0.012 emu/g at $H^{*}/T = 2.5$. 
Since the saturation magnetization is $N_{g}\mu_{B}$ , the value of $N_{g}$
can be estimated as $N_{g} = 1.3 \times 10^{18}$/g.  We also note $N_{g} = 4.7
\times 10^{18}$/g ($N_{g}\mu_{B} = 0.044$ emu/g) in the limit of $H^{*}/T
\rightarrow 0$, which is evaluated from the Curie-Weiss constant (see
Sec.~\ref{resultF}).  Alternative method to determine $N_{g}$ is as
follows.  We assume that $M_{ZFC}= \chi_{d}H + N_{g}\mu_{B} \tanh(\beta
\mu_{B}H)$.  The magnetization $M_{ZFC}$ at fixed $T_{1}$ has a local
maximum at $H^{*} = H_{1}^{*}$: $N_{g} = -1.6053 \times 10^{24}$ $T\chi_{d}
\cosh^{2}(0.671713 H_{1}^{*}/T_{1})$.  Using the values of $H_{1}^{*}$ and
$T_{1}$ determined from the inset of Fig.~\ref{fig:eleven} and $\chi_{d} =
-1.44 \times 10^{-7}$ emu/g, $N_{g}$ is calculated as $N_{g} = (0.91 \pm
0.05) \times 10^{18}$/g.  This value of $N_{g}$ is close to that for
1T-TaS$_{2}$ ($N_{g} = 1.3 \times 10^{18}$/g) which is calculated from the
Curie-Weiss constant ($C_{g} = 0.806 \times 10^{-6}$ emu/g) obtained by
DiSalvo and Waszczak.\cite{DiSalvo1980} 
For comparison, in Fig.~\ref{fig:twelve} we also show the
Brillouin function with these values of $N_{g}$, as a function of
$H^{*}/T$.  We do not find any reasonable value of $N_{g}$, which leads
to good agreement between the result and the Brillouin function over the
whole range of $H^{*}/T$.  Similar behavior has been also observed in
1T-TaS$_{2}$: the downward deviation of the observed magnetization from the
Brillouin function (corresponding to the case of $N_{g} = 2.1 \times
10^{18}$/g in Fig.~\ref{fig:twelve}) occurs for $H^{*}/T \geq
0.25$.\cite{Inada1981} One of the reason for such a difference between the
theory and experiment is that that $U$ is assumed to be constant in the
above model.  The intra-state Coulomb energy $U_{i} = U(\epsilon_{i})$ is
considered to decrease with increasing $\epsilon_{i}$.  The total energy
$2\epsilon_{i} + U_{i}$ is needed for the state i to be occupied by the
spin-up and spin-down electrons.  If $2\epsilon_{i} + U_{i} > 0$, the
$i$-th state is occupied by a single electron which behaves as a free spin. 
If $2\epsilon_{i} + U_{i} < 0$, the $i$-th state is occupied by paired
electrons.\cite{Ikehata1981} The number of free spins is given by
$N(0)\Delta U$, where $\Delta U$ is the region where free spins can situate
on the localized states.  The replacement of $U$ by $\Delta U$ ($< U$)
leads to a decrease in the saturation magnetization.  Thus the downward
deviation of the magnetization from the Brillouin function with $U$ in
Ta$_{2}$S$_{2}$C and 1T-TaS$_{2}$ can be qualitatively explained in terms
of this replacement.

\section{CONCLUSION}
Two phenomena (the superconductivity and the Anderson localization effect) are
observed in Ta$_{2}$S$_{2}$C. The structure of Ta$_{2}$S$_{2}$C can be
viewed as a sum of Ta-C layers and TaS$_{2}$-type structure.  The
superconductivity is mainly due to the Ta-C layers, while the
Anderson localization effect is due to TaS$_{2}$-type structure. 
Ta$_{2}$S$_{2}$C undergoes successive superconducting transitions of a
hierachical nature at $T_{cl} = 3.61 \pm 0.01$ K and $T_{cu} = 9.0 \pm 0.2$
K. The intermediate phase between $T_{cu}$ and $T_{cl}$ is a intra-grain
superconductive state occurring in the Ta-C layers in
Ta$_{2}$S$_{2}$C, while the low temperature phase below $T_{cl}$ is a
inter-grain superconductive state.  The $T$ dependence of magnetic
susceptibility at $H$ well above $H_{c2}^{(l)}(0)$ and the $H$ dependence
of $M_{ZFC}$ at low $T$ and high $H$ are described by a sum of a
diamagnetic susceptibility and a Curie-like behavior.  The latter shows
that the Anderson localization effect occurs in the 1T-TaS$_{2}$-type
structure in Ta$_{2}$S$_{2}$C, leading to the localized magnetic moments due
to unpaired electrons just below $\epsilon_{F}$.

\begin{acknowledgments}
The authors are grateful to Pablo Wally, Technical University of Vienna,
Austria (now Littlefuse, Yokohama, Japan) for providing them with the
samples.  One of the authors (J.W.) acknowledges financial support from the
Ministry of Cultural Affairs, Education and Sport, Japan, under the grant
for young scientists no.  70314375 and from Kansai Invention Center, Kyoto,
Japan.
\end{acknowledgments}

\end{document}